\documentclass[%
 reprint,
superscriptaddress,
 amsmath,amssymb,floatfix,
]{revtex4-2}

\usepackage{graphicx}% Include figure files
\usepackage{dcolumn}% Align table columns on decimal point
\usepackage{comment}
\usepackage{bm}% bold math
\usepackage{color}

\begin{document}

\title{Rheology of dense vibrated granular flows:\\ non-monotonic response controlled by granular temperature}% Force line breaks with \\
\author{A. Plati}
\affiliation{Universit\'e Paris-Saclay, CNRS, Laboratoire de Physique des Solides, 91405 Orsay, France}
\author{G. Petrillo}
\affiliation{The Earth Observatory of Singapore, Nanyang Technological University, Singapore}
\author{L. de Arcangelis}
\affiliation {Department of Mathematics and Physics, University of Campania ``Luigi Vanvitelli", 81100 Caserta, Italy}
\author{A. Gnoli}
\author{A. Puglisi}
\affiliation{Institute for Complex Systems CNR, P.le Aldo Moro 2, 00185, Rome, Italy}
\affiliation{Department of Physics, University of Rome “La Sapienza”, P.le Aldo Moro 5, 00185 Rome, Italy}
\affiliation{INFN, Sezione Roma2, Via della Ricerca Scientifica 1, 00133, Rome, Italy}
\author{A. Sarracino}
\affiliation {Department of Engineering, University of Campania ``Luigi Vanvitelli", 81031 Aversa (CE), Italy}
\author{E. Lippiello}
\affiliation {Department of Mathematics and Physics, University of Campania ``Luigi Vanvitelli", 81100 Caserta, Italy}

\date{\today}

\begin{abstract}
We study the rheology of dense granular materials subjected to vertical vibration {by} using numerical simulations of a stress-imposed vane rheometer. The effective viscosity increases with confining pressure, decreases with vibration amplitude, and exhibits a non-monotonic dependence on frequency: weakening is observed at intermediate frequencies but is lost at high frequencies. We show that the rheological response is governed by the balance between grain-scale agitation energy and the stabilizing effect of confinement. This framework reconciles previously observed trends in viscosity and friction weakening and emphasizes the central role of energy injection and dissipation in determining granular flow properties under vibration.\end{abstract}

\maketitle

\section{Introduction}
 
Granular materials are ubiquitous in natural environment and applications~\cite{andreotti2013granular,puglisi2014transport}. Depending on the external forcing, they can exhibit both solid-like or fluid-like behaviour~\cite{jaeger1996granular,eshuis2007phase}. The rheology of granular materials in shear cells or flowing over inclined planes has been extensively investigated~\cite{bagnold1954experiments,jop2006constitutive,forterre2008flows,da2005rheophysics,bouzid2013nonlocal,fall2015dry,deboeuf2023cohesion}. At the same time, the role of vibration in determining granular flow properties has been highlighted in several theoretical, numerical and experimental studies~\cite{melosh1996dynamical,dijksman2011jamming,hanotin2012vibration,wortel2014rheology,giacco2015dynamic,gnoli2016unified,gnoli2018controlled,leopoldes2020triggering,Plati2021Getting,Clark2023,irmer2025granular,d2025rheological}. External vibrations usually increase the flowability of a granular system and this can have dramatic consequences as in the case of earthquake-triggered landslides~\cite{garcia2010assessment,leopoldes2020triggering} or can be exploited to facilitate the manipulation of granular systems in applications~\cite{wang2021investigation}. In laboratory experiments, vertical vibration is known to reduce to zero the yield stress and produce non-monotonic flow curves~\cite{dijksman2011jamming}. It may also induce a transition from thinning to collisional thickening~\cite{gnoli2016unified}.

The general setting for experiments or simulations to probe the rheology of granular materials under sinusoidal vibration considers the role of three main parameters: the vibration amplitude $A$, the vibration frequency $f$ and the confining pressure $p$. Depending on the specific setup, different quantities can be used as proxies of the ``resistance to flow" of the granular system. When the shear rate $\dot{\gamma}$ is imposed, one usually measures the macroscopic friction coefficient $\mu=\tau/p$ where  $\tau$  is the resulting shear stress. 
Conversely, if the shear stress is imposed, the flow is characterized by the effective viscosity $\eta=\tau/\dot{\gamma}$. Non-vibrated granular systems have been extensively studied by focusing on the so-called flow curves $\mu(\dot{\gamma})$ and $\eta(\tau)$~\cite{fall2015dry,forterre2008flows,DaCruz2002}, however - when vibrations are taken into account - the goal is to characterize the flowability of the material either through $\mu(\dot{\gamma},A,f,p)$ or $\eta(\tau,A,f,p)$. 
To restrict the parameter space, it is sometime useful to consider $\dot{\gamma}$ and $\tau$ fixed, focusing on $\mu(A,f,p)$ and $\eta(A,f,p)$. 
In recent years, many experimental and numerical observations on these quantities have been reported~\cite{gnoli2018controlled,Plati2021Getting,Clark2023,irmer2025granular}.

\begin{figure}
\includegraphics[width=0.99\columnwidth]{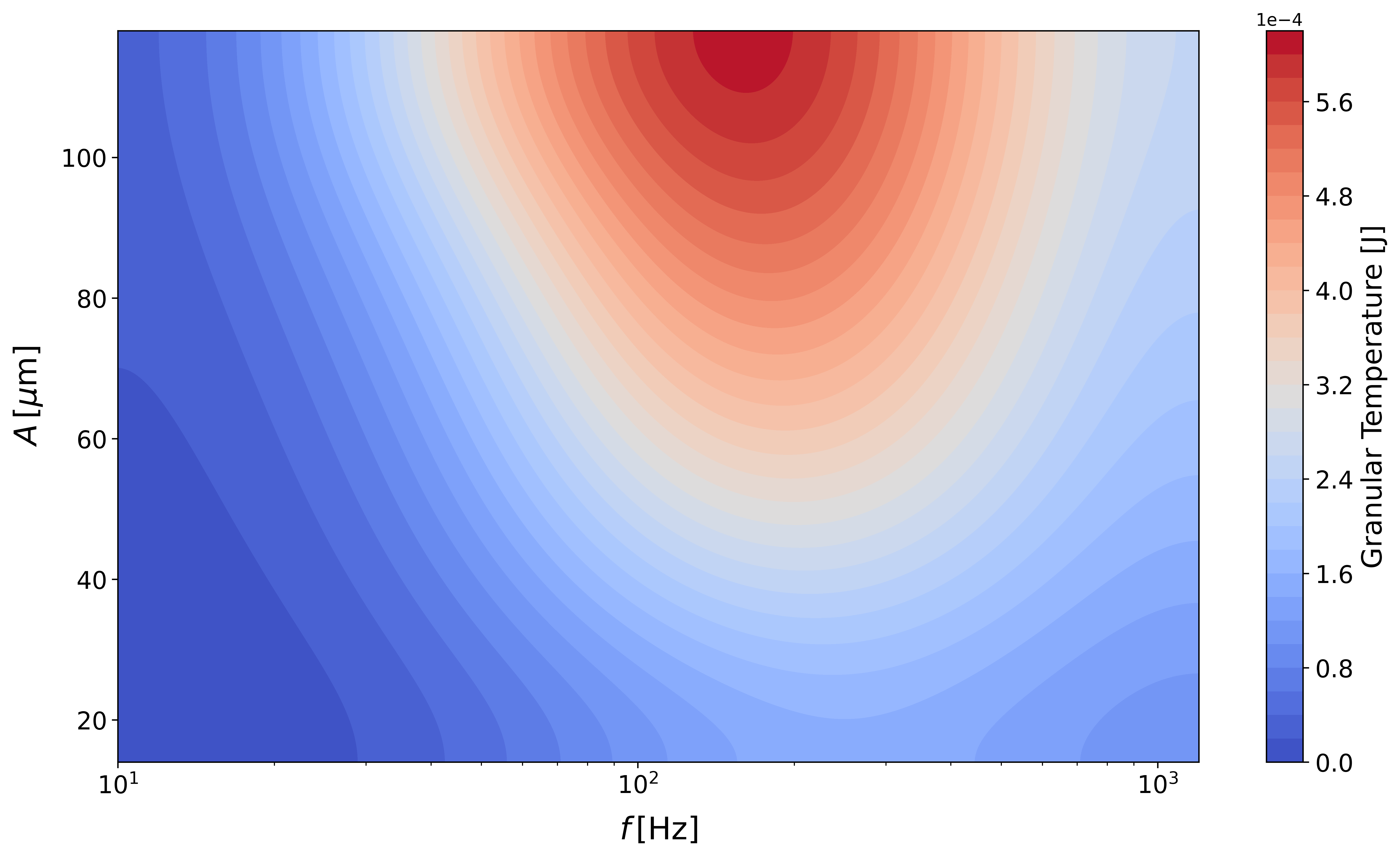}
    \centering
    \caption{ Granular temperature as a function of the oscillation amplitude $A$ and frequency $f$. The granular temperature increases monotonically as a function of $A$ at any fixed frequency. Meanwhile, for any fixed value of $A$, it exhibits a non-monotonic behavior, reaching a maximum at an intermediate frequency. This plot is obtained by extending the numerical results of Ref.~\cite{Plati2021Getting} to a broader range of $A$ and $f$ values.
    } 
    \label{fig:PhaseDiag}
\end{figure}

In Ref.~\cite{gnoli2018controlled}, the effective viscosity of a dense, weakly fluidized granular system was investigated experimentally and numerically by varying the driving amplitude $A$ and  frequency $f$ under a fixed confining load. 
It was found that $\eta$ decreases monotonically with $A$, whereas its dependence on $f$ is non-monotonic: the viscosity decreases until a minimum at a frequency $f^*$ that shifts to lower values as the dissipative properties of the medium increase. 
A subsequent work~\cite{Plati2021Getting} showed that these trends are linked to the behavior of the average kinetic energy of the system (i.e. the granular temperature, see Sec.~\ref{sec:effvisc}A for its definition), which exhibits a maximum at $f^*$. 
In Fig.~\ref{fig:PhaseDiag}, we show the granular temperature as a function of $f$ and $A$ obtained {by} extending the numerical dataset of Ref.~\cite{Plati2021Getting}. It increases monotonically with $A$ when the frequency is fixed and exhibits a non-monotonic trend with $f$ when the amplitude is fixed, which correlates with the behavior of the effective viscosity obtained in experiments~\cite{gnoli2018controlled}.
These results suggest that rheology is primarily controlled by the efficiency with which the medium absorbs energy from external vibration. 
As discussed in~\cite{Plati2021Getting}, while the non-monotonic dependence on $f$ resembles a resonance phenomenon, the fact that $f^*$ depends on dissipation rules out a standard resonance, in which the peak position is fixed by the elastic properties of the system. 
The proposed interpretation instead attributes the non-monotonicity to the competition between the energy injected by the vibration mechanism, and the dissipation rate of the medium (which is related to the grain-grain collision rate). Indeed, both these competing effects are enhanced by the driving frequency, producing a non-monotonic trend. 

 More recently, Clark et al.~\cite{Clark2023} examined the friction coefficient $\mu(A,f,p)$ and observed a similar non-monotonic trend with frequency. However, rather than focusing on the frequency dependence, they analysed the behavior of the minimum friction coefficient $\mu_{\min}(A,p)$ which was found to be controlled by the variable $A^2/p$. 
This scaling was motivated by the idea that friction weakening requires vibration amplitudes large enough to disrupt the underlying force network: while shaking at sufficiently high acceleration %overcoming gravity 
is enough to break individual contacts, an additional threshold in $A^2/p$ is necessary to destabilize mesoscale structures such as force chains. 

In our study we go beyond this interpretation, based on the evidence that,
regardless of whether the thresholds in shaking acceleration and $A^2/p$ are exceeded, friction weakening is always lost at sufficiently high $f$. Following the interpretation of~\cite{Plati2021Getting}, we highlight that the driving frequency governs the efficiency of energy transfer into the granular medium, which in turn controls the system's rheological properties.  
The centrality of energy transfer has also been evidenced in more recent studies, focusing on local rheological responses~\cite{irmer2025granular}.
We therefore emphasize that any comprehensive description of friction weakening must include a detailed analysis of a wider frequency range. 

In this article, we build on these findings to provide a unified framework for understanding the interplay between dissipation, energy injection, and rheological response in realistic vibrated granular systems. In Sec.~\ref{sec:effvisc} we describe our numerical setup which is a realistic \textit{in silico}  reproduction of the experimental system for vane rheometry studied in \cite{gnoli2018controlled,Plati2021Getting}. In the same section we also provide an overview of the rheological response for different shaking parameters and confining pressures. This is done by discussing the behavior of the effective viscosity $\eta(A,f,p)$. This analysis is complemented by the introduction of a minimal model (detailed in Appendix~\ref{app:model}), which accounts for the combined effect of shaking amplitude and confining pressure. In Sec.~\ref{sec:unified}, we show that the key variable controlling $\eta$ involves the ratio between the average kinetic energy of the grains and the confining pressure. There we also discuss a physical interpretation of these results based on the competition between two energy scales, the first related to grain motion and second related to the stability of local structures. In Sec.~\ref{sec:conc}, we conclude our paper by discussing the link between our results and previous studies; we also provide a brief outlook for future developments of this work. 

\begin{figure}
\includegraphics[width=0.7\columnwidth]{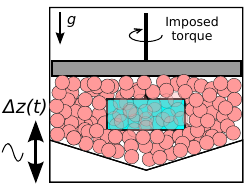}
    \centering
    \caption{{Sketch of the experimental setup used in Ref.~\cite{gnoli2018controlled,Plati2021Getting}. A rotating vane with an imposed constant torque is immersed in a granular medium confined in a conical-shaped container covered by a circular lid. Our numerical setup consists of the \textit{in silico }reconstruction of this apparatus (details in the text).}}
    \label{fig:numSketch}
\end{figure}

\begin{figure*}
\includegraphics[width=0.99\textwidth]{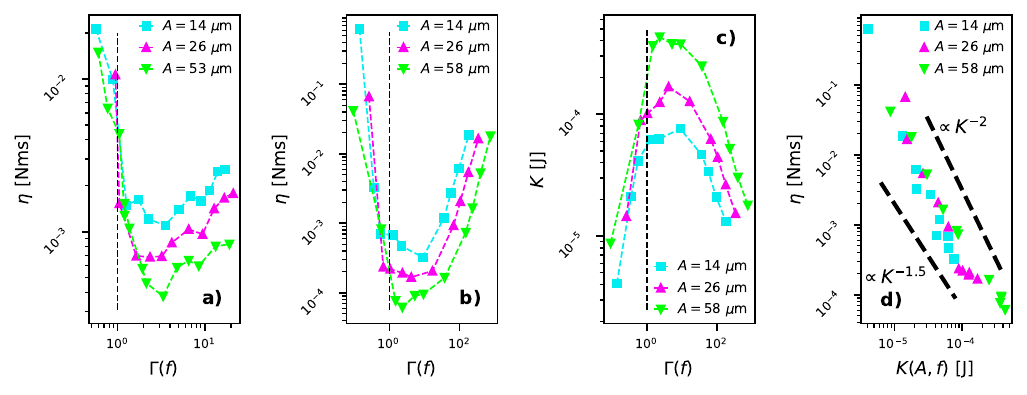}
    \centering
    \caption{ a) Effective viscosity as a function of the rescaled acceleration (varied through $f$)  for three shaking amplitudes. Experimental data from~\cite{gnoli2018controlled}.  b) Analogous results by numerical simulations. Each point is obtained by averaging over 3 statistically independent runs. c) Kinetic energy of the granular medium as a function of $\Gamma$ (varied through $f$) for three shaking amplitudes (simulations performed without the rotating vane). d) Effective viscosity as a function of the granular temperature in numerical simulations. 
    The decreasing trend is well approximated by the scaling $\eta \propto K^\alpha$ with {$\alpha\in[-2,-1.5]$}. {Effective viscosities are obtained from angular velocities expressed in degrees per seconds.}
    }
    \label{fig:ExpAndSim}
\end{figure*}

\section{Effective viscosity of the granular medium}\label{sec:effvisc}

Our numerical setup is the \textit{in silico}  reconstruction of the experimental apparatus used in~\cite{gnoli2018controlled,Plati2021Getting} and it has been shown to provide results in agreement with experiments in previous studies~\cite{plati2019dynamical,Plati2021Getting} {(see Fig.~\ref{fig:numSketch})}.
We use the Discrete Element Method (DEM)~\cite{Cundall79} to perform numerical simulations of a monodispersed granular system with $N=2600$ particles of radius $r=2$ mm and mass $m=0.27$ g confined in a cylinder with a conical-shaped base and covered by a circular lid of mass $M$ and radius $R=45$ mm, which matches the area of the cylinder's round face. The whole system feels gravity acceleration $g$. A rectangular-shaped vane, able to rotate around a fixed vertical axis, is immersed in the granular medium and subjected to a constant torque $\mathcal{T}=0.006$ Nm. The granular material is subjected to a sinusoidal vertical displacement $\Delta z(t)=A\sin(2\pi f t)$ which is imposed on the system's container. The torque value is chosen so that the vane does not move in the absence of vibration. We implement our numerical simulations through the LAMMPS package \cite{Plimpton1995,Plimpton2022,LammpsSiteGran} which uses the Hertz-Mindlin contact model \cite{Zhang2005,LammpsSiteGran} to describe grain-grain, grain-vane, grain-lid and grain-wall interactions. More technical details about our numerical methods are provided in Appendix A.

\subsection{Non-monotonic behavior as a function of the shaking frequency} \label{sec:effvisc_nonMon}
{By} following a microrheology approach, we measure the angular velocity $\Omega(t)$ of the vane  {(always expressed in degrees per seconds)} as a function of time and explore its behaviour {by} varying the frequency $f$ and the amplitude $A$ of the imposed shaking. In view of this, we introduce the rescaled acceleration $\Gamma=(2\pi f)^2A/g$.

 We have checked that, for all the explored combinations of $A$ and $f$, our system always lies in a weakly fluidized regime which means that grains move inside the cage formed by the surrounding particles without experiencing substantial rearrangement except for that induced by the vane's motion. 
Once the vibration and the torque are switched on, after a $t_0$-long initial transient, the vane reaches a stationary state with an average velocity $\langle \Omega \rangle = 1/T \int_{t_0}^{t_0+T}\Omega(t)dt$ which can be thought as a proxy of the shear rate. For all our simulations we have $t_0=13$ s and $T=33$ s. We define the effective viscosity of the granular medium as $\eta=\mathcal{T}/|\langle \Omega \rangle|$. The choice of this variable is motivated by our previous experiments~\cite{gnoli2018controlled} showing a non-monotonic behaviour of $\langle \Omega \rangle$ and in turn of $\eta$ as a function of the frequency. In Fig.~\ref{fig:ExpAndSim}a-b we provide a comparison between experimental and numerical data exhibiting the resulting non-monotonic dependence of $\eta$ as a function of $\Gamma$ (varied through $f$) for different $A$.
%This dependence of $\langle\Omega \rangle$ (and in turn of $\eta$) on the frequency has been already investigated in \cite{Plati2021Getting}. A
As highlighted by the vertical lines in the plots, the decreasing branch of the $\eta(\Gamma)$ curves can be explained in terms of the fluidization threshold around $\Gamma=1$~\cite{gnoli2018controlled}, where the shaking acceleration becomes equal to $g$. The physical mechanism underlying the viscosity strengthening at larger $\Gamma$ has been discussed in a previous work~\cite{Plati2021Getting}, where we showed that it is linked to the behavior of the average kinetic energy of the granular medium in the steady state $K= 1/T \int_{t_0}^{t_0+T} K_{\text{inst}}(t)dt$, where $K_{\text{inst}}(t)=\frac{m}{2N}\sum_iv_i^2(t)$ and $v_i(t)$ is the instantaneous speed of the grain $i$. The variable $K$ is usually called granular temperature (or directly proportional to it through the number of degrees of freedom which is constant here) and displays itself a non-monotonic dependence on the vibration frequency as shown in Fig.~\ref{fig:ExpAndSim}c. This phenomenon originates from the competing effects of injected vibration energy and (inelastic) grain-grain collision rate which both increase with $f$~\cite{Plati2021Getting}. A more detailed numerical analysis supporting this interpretation based on energy dissipation can be found in Appendix~\ref{app:ediss}.  The behavior of $\eta$ as a function of the driving parameters $A$ and $f$ is then effectively controlled by $K$, as highlighted in Fig.~\ref{fig:ExpAndSim}d. 
{The decreasing trend is well approximated by the power-law scaling $\eta \propto K^\alpha$, with $\alpha\in[-2,-1.5]$ (see the guides for the eye).}
We finally point out here that to investigate the effect of the granular temperature on $\eta$ without considering trivial correlations due to the blade motion, we measured $K$ in simulations \emph{without} the rotating blade. 

\subsection{Effect of normal pressure}\label{sec:effvisc_press}
\begin{figure*}
\includegraphics[width=0.99\textwidth]{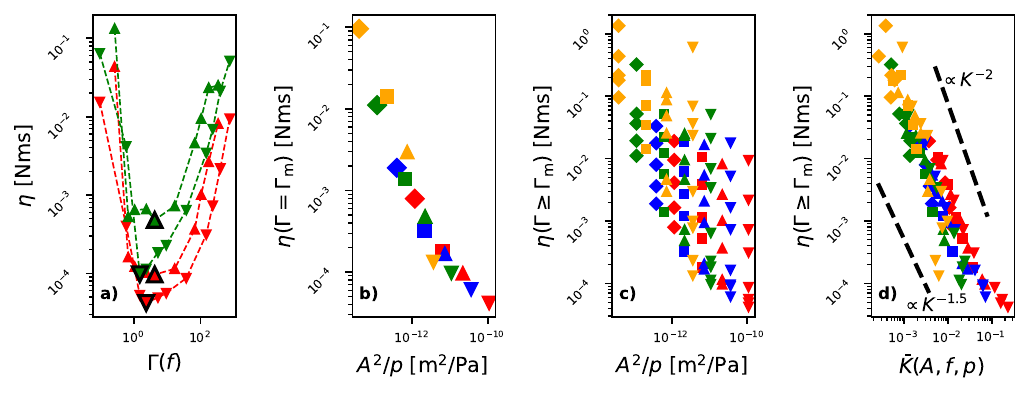}
    \centering
    \caption{ a) Effective viscosity as a function of $\Gamma$ (varied through $f$) for two couples of amplitudes and pressures. b)  Minimum effective viscosity as a function of $A^2/p$ for four pressures. c) Effective viscosity in the fluidized state ($\Gamma > \Gamma_{\text{m}}$) as a function of $A^2/p$ for four pressures and amplitudes.  d) Effective viscosity in the fluidized state ($\Gamma > \Gamma_{\text{m}}$) as a function of $\bar{K}\propto K/p$ for four pressures and amplitudes. Diamond, square, upper- and lower-triangle markers correspond to $A=6,14,26,58$ $\mu$m respectively. Red, blue, green, yellow markers correspond to $p=31,92,308,924$ Pa respectively. {Effective viscosities are obtained from angular velocities expressed in degrees per seconds.}   
    } 
    \label{fig:EffPressFinScal}
\end{figure*}

We now focus on the behaviour of $\eta$ as a function of $\Gamma$ (varied through $f$) for different couples of $\{A, p\}$ where $p=M g/(\pi R^2)$ is the normal pressure due to the lid's weight. In Fig.~\ref{fig:EffPressFinScal}a, we observe that all curves decrease up to the fluidization threshold and then reach a minimum $\eta_{\text{min}}$ (highlighted with a larger mark) at $\Gamma=\Gamma_{\text{m}}$ before showing a gradual increase.  As expected, increasing $A$ reduces the effective viscosity while raising $p$ has the opposite effect. In order to first focus on the most fluidized regime, we get rid for the moment of the $\Gamma$ dependence {and consider} the behavior of $\eta_{\text{min}}(A,p)$.
In Fig.~\ref{fig:EffPressFinScal}b, we show that $\eta_{\text{min}}(A,p)$ is mainly controlled by the variable $A^2/p$. 
Remarkably, a similar data collapse has been previously observed also for the macroscopic friction coefficient in simulations of 2D sheared granular materials under vertical vibrations \cite{Clark2023}. Our results generalize this observation to the effective viscosity of a  3D granular system probed through stress-imposed microrheology. 
The role of the variable $A^2/p$ was explained in  Ref.~\cite{Clark2023} based on physical mechanisms involving the rupture of force chains in the bulk of the granular medium.
In Appendix~\ref{app:model} we discuss a simple mechanical model which is able to reproduce the $A^2/p$ scaling and allows us to trace it back  to the interplay between energy injection and non-linear elastic deformation. This shows that force chain breaking is not necessary to explain the observed scaling.

\section{A unified Picture}\label{sec:unified}

Now we point out that the observation of a non-monotonic behavior of $\eta$ as a function of the shaking frequency (Sec.~\ref{sec:effvisc_nonMon}) is in striking contrast with the idea that the rheological response of the granular medium is mainly controlled by $A^2/p$. While the viscosity weakening branch can be understood by combining the acceleration threshold $\Gamma=1$ and the $A^2/p$ scaling, the raising branch $\eta(\Gamma\ge\Gamma_\text{m})$ is not.  Indeed, simulations performed at $\Gamma\ge\Gamma_\text{m}$ (varied through $f$), with the same amplitude and pressure can manifest effective viscosities that vary over different orders of magnitude, see Fig.~\ref{fig:EffPressFinScal}c. {At the same time, we verified that the system's kinetic energy is only marginally influenced by $p$ (see Appendix~\ref{sec:appA})}, thus the effect of $p$ on the effective viscosity (see Fig.~\ref{fig:EffPressFinScal}a) can't be explained with the dependence of $K$ on $p$. Based on our numerical analysis, we define an energy scale related to the confining pressure $K_p=p\pi r^3$ (i.e. the work necessary to move of a distance $r$ against a constant force $p\pi r^2$) and propose the energy ratio $\bar{K}=K/K_p$ as the main variable controlling the effective viscosity of the system. In Fig.~\ref{fig:EffPressFinScal}d, we plot $\eta(\Gamma\ge\Gamma_\text{m})$ obtained for all the compatible frequencies, amplitudes and pressures varied in our analysis, showing a good collapse on the master curve $\eta\propto \bar{K}^\alpha$. {Also in this case, our data are well described by a power law $\propto \bar{K}^{\alpha}$, with $\alpha\in[-2,-1.5]$.}

The adimensionalized energy $\bar{K}$ can be interpreted as a competition between the typical agitation energy of the grains inside their cage $K$ and the typical energy (set by the confining pressure) needed to escape from such a cage $K_p$. As visible from the $x$-axis of Fig.~\ref{fig:EffPressFinScal}d, our simulations are always in the regime $\bar{K}\ll 1$ where we expect spontaneous grain rearrangement to be extremely rare. The main idea behind a $\eta (\bar{K})$ dependence is that, even in this weakly fluidized regime, the larger $\bar{K}$, the easier it will be for the rotating blade to break grain cages in order to move. 

We point out that the ratio of energies defining $\bar{K}$ reminds of the competition between thermal agitation and activation energy barriers that characterizes the viscosity divergence in glass-forming supercooled liquids~\cite{debenedetti2001supercooled} with, however, the important difference that in these systems viscosity divergences are usually exponential or even steeper while in our case we observe a power law.
This qualitative analogy with thermal systems is nevertheless conceptually useful to understand the relationship between rheological response and $\bar{K}(A,f,p)$ in our granular setup. 
A peculiarity of thermal systems is that temperature can be directly controlled by equilibrating the system with a thermal bath. Then, as a consequence of thermodynamic equilibrium, also the average kinetic energy of the system is fixed and proportional to the temperature.  Instead, in a granular system, the average kinetic energy $K(A,f)$ depends non-trivially on the driving parameters. In other words, the relationship between injected and adsorbed energy is much more complex when thermodynamic equilibrium is violated, as in granular systems. Therefore, to understand how the driving parameters $A$ and $f$ control the rheology of a granular system it is important to first clarify how they act on its granular temperature $K$. Our numerical analysis shows that, since $\eta$ depends non-monotonously on $f$, there is no simple scaling in the form $\propto A^\beta f^\lambda$, with $\beta$ and $\lambda$ arbitrary exponents, that could predict the behavior of the effective viscosity. Panel c and d of Fig.~\ref{fig:EffPressFinScal} show in fact that $\eta$ depends on $A$ and $f$ through $K(A,f)$, which embodies the non-monotonic dependence on $f$.

\section{Conclusions and link with previous studies}\label{sec:conc}
In this paper we investigated the rheology of a dense vibrated granular system by means of realistic numerical DEM simulations of a stress imposed vane rheometry setup. We varied the vibration amplitude $A$ and frequency $f$ of the vertical sinusoidal shaking as well as the confining pressure $p$. We found that the variable controlling the effective viscosity is $\bar{K}=K/K_p$ where $K(A,f)$ is the typical agitation energy of a grain in the cage formed by the surrounding particles and $K_p \propto p$ is an energy scale set by the confining pressure that could be interpreted as the energy needed to escape from such a cage. Our proposed scaling takes into account the non-monotonic effect of $f$ on $\eta$ through the energy transfer mechanism discussed in Ref.~\cite{Plati2021Getting} and Appendix~\ref{app:ediss}.

We now discuss the connection between our study and previous works on the rheology of vibrated granular materials in simplified geometries. We start {by} referring to the numerical study reported in~\cite{Clark2023} which focuses on the macroscopic friction coefficient $\mu$ of a granular system subjected to a vertical sinusoidal shaking (with frequency $f$ and amplitude $A$) in a double plate geometry under a confining pressure $p$. Remarkably, similarly to the effective viscosity of our system, the frequency dependence of the friction coefficient $\mu(f)$ reported in this study exhibits a non-monotonic trend. When considering the full $\mu(A,f,p)$ dependence, the authors showed that the friction weakening behavior can be predicted by considering the region of control parameters identified by $\Gamma>1$ and $A^2/p>a$ where $a$ is a threshold that depends on the restitution coefficient in the grain-grain interaction. However, the provided criterion only holds in the frequency domain where $\mu(f)$ decreases, thus not including in the analysis its non-monotonic behavior.    
We want to stress here that our study marks a step forward with respect to Ref.~\cite{Clark2023} for two key points. First, when the frequency dependence is not taken into account, we showed that the variable  $A^2/p$ proposed for the friction-weakening behavior also controls the effective viscosity in our more realistic setup. Moreover, we have also proposed a minimal model for such a scaling in Appendix \ref{app:model}. Second, when the full frequency dependence is taken into account, we showed that it is not possible to predict the viscosity reduction by simply considering some thresholds on $A^2/p$ and $\Gamma$ because  it depends on the control parameters through $K(A,f)$, which is the key variable embodying the non-monotonic dependence on $f$. We believe that the same idea could be used to generalize the analysis on the friction coefficient done in Ref.~\cite{Clark2023} by taking into account the non-monotonic $\mu(f)$ dependence. It is also important to point out that the relationship between granular temperature and rheology of vibrated granular materials was already introduced (although with a less systematic analysis) in our previous study~\cite{Plati2021Getting} and later in Ref.~\cite{irmer2025granular} by Irmer et al. that showed how the granular temperature controls the local rheology in a double plate geometry. 
However, we point out that the latter study has been performed in a more idealized numerical setup without a direct experimental counterpart and without any specific mention of non-monotonic behaviors of rheological observables and granular temperature. 

{For future studies, it would be useful to develop a theoretical framework to estimate the exponent $\alpha$ of the power-law that links $\eta$ and $\bar{K}$.}
Moreover, inspired by seminal studies on non-vibrated granular flows~\cite{forterre2008flows}, it would  be interesting to study the combined effect of vibration driving parameters and confining pressure also on the volume fraction of vibrated granular systems.

\begin{acknowledgments}

A. Plati acknowledges funding from the Agence Nationale
de la Recherche (ANR), France, grant ANR-21-CE06-0039. AG, AP and AS  acknowledge funding from the Italian Ministero dell’Università e della Ricerca under the programme PRIN 2022 ("re-ranking of the final lists"), number 2022KWTEB7, cup B53C24006470006. EL acknowledges funding from the Italian Ministero dell’Università e della Ricerca under the programme PRIN 2022 PNRR P202247YKL, CUP B53D23033430001.
\end{acknowledgments}

\section*{Data Availability}
The data that support the findings of this article are openly available~\cite{zenodo}.

\bibliography{biblioGrRh}

@article{Plati2021Getting,
  title = {Getting hotter by heating less: How driven granular materials dissipate energy in excess},
  author = {Plati, A. and de Arcangelis, L. and Gnoli, A. and Lippiello, E. and Puglisi, A. and Sarracino, A.},
  journal = {Phys. Rev. Research},
  volume = {3},
  issue = {1},
  pages = {013011},
  numpages = {12},
  year = {2021},
  month = {Jan},
  publisher = {American Physical Society},
  doi = {10.1103/PhysRevResearch.3.013011},
  url = {https://link.aps.org/doi/10.1103/PhysRevResearch.3.013011}
}

@article{hanotin2012vibration,
  title={Vibration-induced liquefaction of granular suspensions},
  author={Hanotin, Caroline and Kiesgen de Richter, S and Marchal, Philippe and Michot, Laurent J and Baravian, Christophe},
  journal={Phys. Rev. Lett.},
  volume={108},
  number={19},
  pages={198301},
  year={2012},
  publisher={APS}
}

@article{jaeger1996granular,
  title={Granular solids, liquids, and gases},
  author={Jaeger, Heinrich M and Nagel, Sidney R and Behringer, Robert P},
  journal={Reviews of modern physics},
  volume={68},
  number={4},
  pages={1259},
  year={1996},
  publisher={APS}
}

@article{eshuis2007phase,
  title={Phase diagram of vertically shaken granular matter},
  author={Eshuis, Peter and Van Der Weele, Ko and Van Der Meer, Devaraj and Bos, Robert and Lohse, Detlef},
  journal={Physics of Fluids},
  volume={19},
  number={12},
  year={2007},
  publisher={AIP Publishing}
}

@article{giacco2015dynamic,
  title={Dynamic weakening by acoustic fluidization during stick-slip motion},
  author={Giacco, Ferdinando and Saggese, Luigi and de Arcangelis, Lucilla and Lippiello, Eugenio and Pica Ciamarra, Massimo},
  journal={Phys. Rev. Lett.},
  volume={115},
  number={12},
  pages={128001},
  year={2015},
  publisher={APS}
}

@article{debenedetti2001supercooled,
  title={Supercooled liquids and the glass transition},
  author={Debenedetti, Pablo G and Stillinger, Frank H},
  journal={Nature},
  volume={410},
  number={6825},
  pages={259--267},
  year={2001},
  publisher={Nature Publishing Group UK London}
}

@article{Clark2023,
  title = {Frictional Weakening of Vibrated Granular Flows},
  author = {Clark, Abram H. and Brodsky, Emily E. and Nasrin, H. John and Taylor, Stephanie E.},
  journal = {Phys. Rev. Lett.},
  volume = {130},
  issue = {11},
  pages = {118201},
  numpages = {6},
  year = {2023},
  month = {Mar},
  publisher = {American Physical Society},
  doi = {10.1103/PhysRevLett.130.118201},
  url = {https://link.aps.org/doi/10.1103/PhysRevLett.130.118201}
}

@article{Cundall79,
author = {Cundall, P. A. and Strack, O. D. L.},
title = {A discrete numerical model for granular assemblies},
journal = {Géotechnique},
volume = {29},
number = {1},
pages = {47-65},
year = {1979},

URL = { 
        https://doi.org/10.1680/geot.1979.29.1.47
    
},
eprint = { 
        https://doi.org/10.1680/geot.1979.29.1.47
    
}

}

@article{melosh1996dynamical,
  title={Dynamical weakening of faults by acoustic fluidization},
  author={Melosh, HJ},
  journal={Nature},
  volume={379},
  number={6566},
  pages={601--606},
  year={1996},
  publisher={Nature Publishing Group UK London}
}

@article{jop2006constitutive,
  title={A constitutive law for dense granular flows},
  author={Jop, Pierre and Forterre, Yo{\"e}l and Pouliquen, Olivier},
  journal={Nature},
  volume={441},
  number={7094},
  pages={727--730},
  year={2006},
  publisher={Nature Publishing Group UK London}
}

@article{bagnold1954experiments,
  title={Experiments on a gravity-free dispersion of large solid spheres in a Newtonian fluid under shear},
  author={Bagnold, Ralph Alger},
  journal={Proceedings of the Royal Society of London. Series A. Mathematical and Physical Sciences},
  volume={225},
  number={1160},
  pages={49--63},
  year={1954},
  publisher={The Royal Society London}
}

@article{da2005rheophysics,
  title={Rheophysics of dense granular materials: Discrete simulation of plane shear flows},
  author={Da Cruz, Fr{\'e}d{\'e}ric and Emam, Sacha and Prochnow, Micha{\"e}l and Roux, Jean-No{\"e}l and Chevoir, Fran{\c{c}}ois},
  journal={Phys. Rev. E},
  volume={72},
  number={2},
  pages={021309},
  year={2005},
  publisher={APS}
}

@article{d2025rheological,
  title={Rheological regimes in agitated granular media under shear},
  author={D’Angelo, Olfa and Sperl, Matthias and Kranz, W Till},
  journal={Phys. Rev. Lett.},
  volume={134},
  number={14},
  pages={148202},
  year={2025},
  publisher={APS}
}

@article{fall2015dry,
  title={Dry granular flows: Rheological measurements of the $\mu$ (I)-rheology},
  author={Fall, Abdoulaye and Ovarlez, Guillaume and Hautemayou, David and M{\'e}zi{\`e}re, C{\'e}dric and Roux, J-N and Chevoir, Fran{\c{c}}ois},
  journal={Journal of rheology},
  volume={59},
  number={4},
  pages={1065--1080},
  year={2015},
  publisher={AIP Publishing}
}

@article{deboeuf2023cohesion,
  title={Cohesion and aggregates in unsaturated wet granular flows down a rough incline},
  author={Deboeuf, Stephanie and Fall, Abdoulaye},
  journal={Journal of Rheology},
  volume={67},
  number={4},
  pages={909--909},
  year={2023},
  publisher={AIP Publishing}
}

@article{Plimpton1995,
title = {Fast Parallel Algorithms for Short-Range Molecular Dynamics},
journal = {Journal of Computational Physics},
volume = {117},
number = {1},
pages = {1-19},
year = {1995},
issn = {0021-9991},
doi = {https://doi.org/10.1006/jcph.1995.1039},
url = {https://www.sciencedirect.com/science/article/pii/S002199918571039X},
author = {Steve Plimpton}
}

@article{plati2019dynamical,
  title={Dynamical collective memory in fluidized granular materials},
  author={Plati, A and Baldassarri, A and Gnoli, A and Gradenigo, G and Puglisi, A},
  journal={Phys. Rev. Lett.},
  volume={123},
  number={3},
  pages={038002},
  year={2019},
  publisher={APS}
}

@article{Plimpton2022,
title = {LAMMPS - a flexible simulation tool for particle-based materials modeling at the atomic, meso, and continuum scales},
journal = {Computer Physics Communications},
volume = {271},
pages = {108171},
year = {2022},
issn = {0010-4655},
doi = {https://doi.org/10.1016/j.cpc.2021.108171},
url = {https://www.sciencedirect.com/science/article/pii/S0010465521002836},
author = {Aidan P. Thompson and H. Metin Aktulga and Richard Berger and Dan S. Bolintineanu and W. Michael Brown and Paul S. Crozier and Pieter J. {in 't Veld} and Axel Kohlmeyer and Stan G. Moore and Trung Dac Nguyen and Ray Shan and Mark J. Stevens and Julien Tranchida and Christian Trott and Steven J. Plimpton}}

@misc{zenodo,

  howpublished = "\url{https://zenodo.org/records/18805723}"
}

@misc{LammpsSiteGran,

  howpublished = "\url{https://docs.lammps.org/pair_granular.html}"
}

@article{DaCruz2002,
  title = {Viscosity bifurcation in granular materials, foams, and emulsions},
  author = {Da Cruz, F. and Chevoir, F. and Bonn, Daniel and Coussot, P.},
  journal = {Phys. Rev. E},
  volume = {66},
  issue = {5},
  pages = {051305},
  numpages = {7},
  year = {2002},
  month = {Nov},
  publisher = {American Physical Society},
  doi = {10.1103/PhysRevE.66.051305},
  url = {https://link.aps.org/doi/10.1103/PhysRevE.66.051305}
}

@book{puglisi2014transport,
  title={Transport and fluctuations in granular fluids: From Boltzmann equation to hydrodynamics, diffusion and motor effects},
  author={Puglisi, Andrea},
  year={2014},
  publisher={Springer}
}

@book{andreotti2013granular,
  title={Granular media: between fluid and solid},
  author={Andreotti, Bruno and Forterre, Yo{\"e}l and Pouliquen, Olivier},
  year={2013},
  publisher={Cambridge University Press}
}

@article{Zhang2005,
  title = {Jamming transition in emulsions and granular materials},
  author = {Zhang, H. P. and Makse, H. A.},
  journal = {Phys. Rev. E},
  volume = {72},
  issue = {1},
  pages = {011301},
  numpages = {12},
  year = {2005},
  month = {Jul},
  publisher = {American Physical Society},
  doi = {10.1103/PhysRevE.72.011301},
  url = {https://link.aps.org/doi/10.1103/PhysRevE.72.011301}
}

@article{gnoli2016unified,
  title={Unified rheology of vibro-fluidized dry granular media: From slow dense flows to fast gas-like regimes},
  author={Gnoli, Andrea and Lasanta, Antonio and Sarracino, Alessandro and Puglisi, Andrea},
  journal={Scientific Reports},
  volume={6},
  number={1},
  pages={38604},
  year={2016},
  publisher={Nature Publishing Group UK London}
}

@article{dijksman2011jamming,
  title={Jamming, yielding, and rheology of weakly vibrated granular media},
  author={Dijksman, Joshua A and Wortel, Geert H and Van Dellen, Louwrens TH and Dauchot, Olivier and Van Hecke, Martin},
  journal={Phys. Rev. Lett},
  volume={107},
  number={10},
  pages={108303},
  year={2011},
  publisher={APS}
}

@article{wortel2014rheology,
  title={Rheology of weakly vibrated granular media},
  author={Wortel, Geert H and Dijksman, Joshua A and Van Hecke, Martin},
  journal={Phys. Rev. E},
  volume={89},
  number={1},
  pages={012202},
  year={2014},
  publisher={APS}
}

@article{irmer2025granular,
  title={Granular temperature controls local rheology of vibrated granular flows},
  author={Irmer, Mitchell G and Brodsky, Emily E and Clark, Abram H},
  journal={Phys. Rev. Lett.},
  volume={134},
  number={4},
  pages={048202},
  year={2025},
  publisher={APS}
}

@article{wang2021investigation,
  title={Investigation on asphalt-screed interaction during pre-compaction: Improving paving effect via numerical simulation},
  author={Wang, Chonghui and Moharekpour, Milad and Liu, Quan and Zhang, Zeyu and Liu, Pengfei and Oeser, Markus},
  journal={Construction and Building Materials},
  volume={289},
  pages={123164},
  year={2021},
  publisher={Elsevier}
}

@article{garcia2010assessment,
  title={Assessment of earthquake-triggered landslide susceptibility in El Salvador based on an Artificial Neural Network model},
  author={Garc{\'\i}a-Rodr{\'\i}guez, MJ and Malpica, JA},
  journal={Natural Hazards and Earth System Sciences},
  volume={10},
  number={6},
  pages={1307--1315},
  year={2010},
  publisher={Copernicus GmbH}
}

@article{gnoli2018controlled,
  title={Controlled viscosity in dense granular materials},
  author={Gnoli, A and De Arcangelis, L and Giacco, F and Lippiello, E and Ciamarra, M Pica and Puglisi, A and Sarracino, A},
  journal={Phys. Rev. Lett},
  volume={120},
  number={13},
  pages={138001},
  year={2018},
  publisher={APS}
}

@article{forterre2008flows,
  title={Flows of dense granular media},
  author={Forterre, Yo{\"e}l and Pouliquen, Olivier},
  journal={Annu. Rev. Fluid Mech.},
  volume={40},
  number={1},
  pages={1--24},
  year={2008},
  publisher={Annual Reviews}
}

@book{Ble00,
  title={Vibrational mechanics: nonlinear dynamic effects, general approach, applications},
  author={Blekhman, Iliya I},
  year={2000},
  publisher={World Scientific}
}

@article{leopoldes2020triggering,
  title={Triggering granular avalanches with ultrasound},
  author={L{\'e}opold{\`e}s, J and Jia, X and Tourin, A and Mangeney, A},
  journal={Phys. Rev. E},
  volume={102},
  number={4},
  pages={042901},
  year={2020},
  publisher={APS}
}

@article{bouzid2013nonlocal,
  title={Nonlocal rheology of granular flows across yield conditions},
  author={Bouzid, Mehdi and Trulsson, Martin and Claudin, Philippe and Cl{\'e}ment, Eric and Andreotti, Bruno},
  journal={Phys. Rev. Lett.},
  volume={111},
  number={23},
  pages={238301},
  year={2013},
  publisher={APS}
}

\clearpage
\appendix
\section{Details on numerical simulations}
\label{sec:appA}
Our DEM simulations make use of the $pair\textunderscore style$ granular command of the LAMMPS package~\cite{Plimpton1995,Plimpton2022,LammpsSiteGran} to implement the Hertz–Mindlin~\cite{Zhang2005,LammpsSiteGran} model for grain–grain, grain–lid, and grain–vane contact dynamics. The lid is composed of 1773 granular particles glued together, and the rotating vane consists of 4 $\times$ 10 particles glued together and overlapping by half a radius. Grain–wall interactions are described by the same model, implemented with the $wall/gran/region$ command. The Hertz–Mindlin model accounts for both the elastic and dissipative responses during mutual compression of the interacting objects and incorporates the dynamics of their relative translational and rotational motions (see~\cite{LammpsSiteGran} for the specific expressions of tangential and normal forces). 

All this requires five parameters to be set, representing the stiffness and viscosity of both the normal ($k_n$, $\eta_n$) and tangential ($k_t$, $\eta_t$) forces, along with the tangential friction coefficient $\mu$. For all our simulations, we fix $k_n = 6.1\times 10^7$ Pa, $\eta_n = 2.9\times 10^7$ (m$\cdot$s)$^{-1}$, $k_t = 7.0\times 10^7$ Pa, $\eta_t = 2.3\times 10^5$ (m$\cdot$s)$^{-1}$, and $\mu = 0.5$ for grain–grain, grain–lid, grain–vane, and grain–wall interactions, which, for simplicity, are considered to be made of the same material. These numerical values are chosen within a range that, in previous studies~\cite{plati2019dynamical,Plati2021Getting}, proved to yield numerical results consistent with experimental observations. As shown in Ref.~\cite{Plati2021Getting}, these parameters influence the energy transfer mechanisms, but the observation of the non-monotonic response as a function of driving frequency remains robust against their variation.

Our simulations are initialized by randomly pouring the grains into the container and then {by} letting the lid fall on them. During the pouring process, we impose a high-amplitude shaking to compact the system more effectively. After pouring, the specific $A$ and $f$ to be probed are set, and the torque on the vane is switched on.

\begin{figure}
\includegraphics[width=0.9\columnwidth,clip=true]{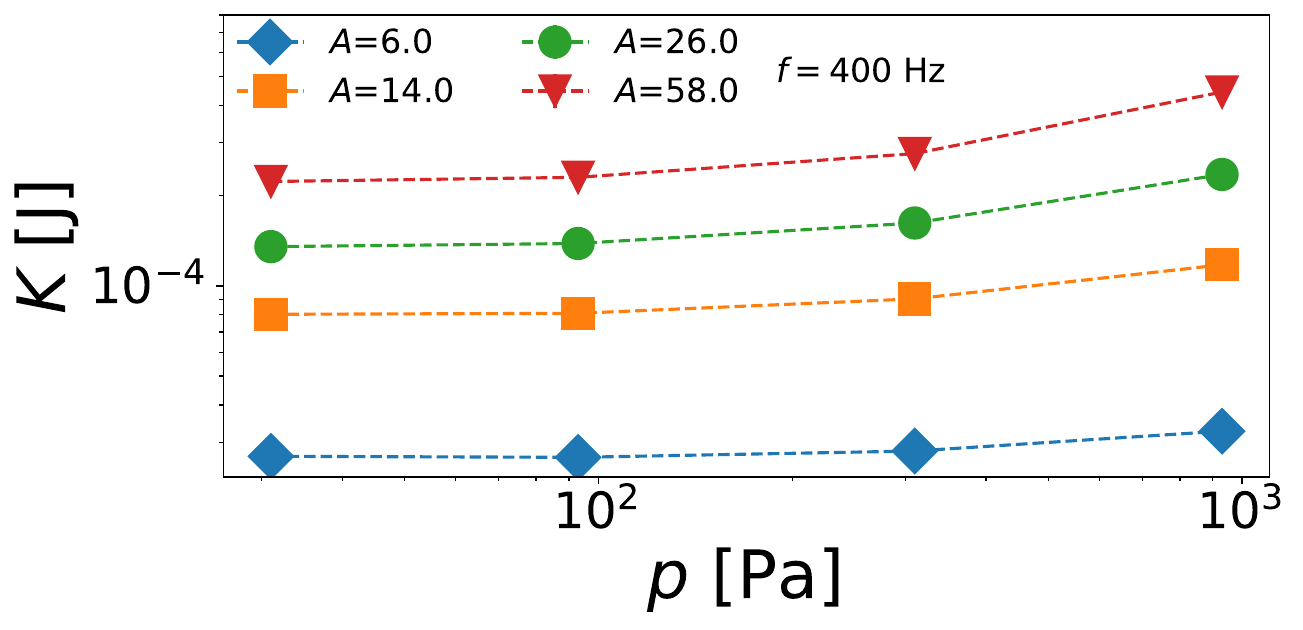}
    \centering
\caption{{Kinetic energy of the granular medium as a function of $p$ for four shaking
amplitudes and $f=400$ Hz.} }
\label{fig:KvspVarioA}
\end{figure}

{To complement the discussion of Sec.~\ref{sec:unified}, we show in Fig.~\ref{fig:KvspVarioA} the plot of the system's kinetic energy as a function of $p$ for different driving amplitudes at a fixed frequency $f=400$ Hz.  We observe that the variation as a function of $p$ over almost two decades is small compared to the one given by varying $A\in[6,58]$ $\mu$m. Comparing this plot with Fig.~\ref{fig:ExpAndSim}c
we do see that such variation is also small compared to the one observed by varying $\Gamma$ through $f$. We don't have a clear explanation for the slight systematic increasing trend of $K$ as function of $p$. One possible reason is that a heavier lid implies a better confinement of the granular medium, thereby enhancing its interaction with the vibrating box and thus improving the energy transfer from the external source. Overall, the main conclusion drawn from this analysis is that the effect of $p$ on $K$ is negligible when compared to the one of $A$ and $f$.}

\section{The role of energy dissipation}\label{app:ediss}
In order to support our interpretation based on energy dissipation for the non-monotonic behaviour of $K$, we studied the behaviour of the average work done by the dissipative forces in our system.  Our analysis starts by considering that in a time interval  $\Delta t$, the variation of the total internal energy in the system (i.e. kinetic plus potiential) is equal to the work $W_{nc}$ done by the non-conservative forces on the system in that interval $U(t+\Delta t)-U(t)=W_{nc}$. Assuming a stationary state and taking the time average of this equality we obtain $\langle W_{nc} \rangle=0$. In other words, different contributions to the work done by non-conservative forces must balance each other.  In the setting used to study the granular temperature of the granular medium (i.e. without the rotating vane), we have the work done by the vibrating container $W_w$ , the work done by the confining lid $W_l$ and finally the work due to the dissipative components in the grain interactions $W_d$ (see Sec.~\ref{sec:appA} and~\cite{LammpsSiteGran}). Thus, assigning the plus sign to the work done \emph{on} the system and the minus sign to the work done \emph{by} the system, we have $W_{nc}=W_w+W_l-W_d$. The average work done by internal dissipative forces can then be expressed as $\langle W_d \rangle=\langle W_w \rangle+\langle W_l \rangle.$ 
\begin{figure}
\includegraphics[width=0.99\columnwidth,clip=true]{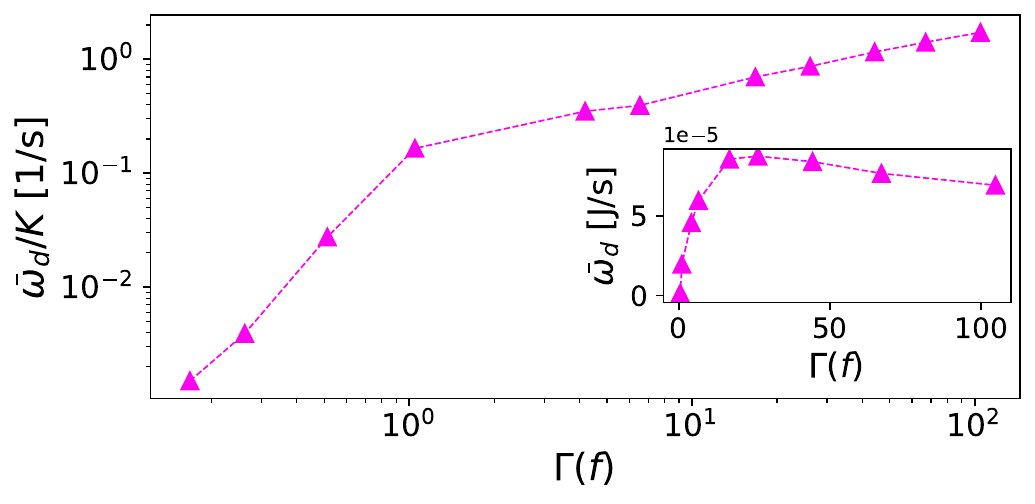}
    \centering
\caption{Ratio of the dissipative power to the granular temperature (main panel) and dissipative power (inset) as a function of $\Gamma$ varied through $f$. Simulations are in the same geometry as the ones discussed in the main text but removing both the vane and the top lid. The driving amplitude is fixed to $A=26$ $\mu$m. }
\label{fig:Ediss}
\end{figure}
In the following, to simplify the analysis, we considered simulations without the lid (i.e. $\langle W_l \rangle=0$ by definition); moreover, stationarity allows us to write down the work contributions during an interval $\Delta t$ in terms of power. All this leads to the central identity for the analysis that follows: 
\begin{equation}
    \omega_d \Delta t = \omega_w \Delta t ,
\label{eq::appRates}
\end{equation}
where $\omega_d$ and $\omega_w$ are, respectively, the power associated to the dissipative interactions and the one associated to the vibrating walls. Practically, from our numerical data, we measured the instantaneous force $\boldsymbol{F}^w_i$ between the walls and each grain $i$ at contact for each simulation step $dt$, along with the displacement $\boldsymbol{dS}_i$  travelled by the grain in the same $dt$. We point out that, to further simplify this numerical measure, we switched off the tangential interactions between the walls and the grains in the simulations performed for this analysis. 
Then, using Eq.~\eqref{eq::appRates} we  obtain the per-particle dissipation power  
\begin{equation}
    \bar{\omega}_d = (Ndt)^{-1} \sum_i \boldsymbol{F}^w_i \cdot \boldsymbol{dS}_i .
\end{equation}
This procedure is much easier than evaluating the dissipation rate {from} the work done by the dissipative contact forces for each couple of interacting grains. Nevertheless, we also performed some auxiliary simulations in a simplified geometry with only a couple of grains to check the consistency of our approach (not shown).

In Fig.~\ref{fig:Ediss}, we plot the ratio  $\bar{\omega}_d/K$ as a function of $\Gamma$ varied through $f$ and observe an increasing monotonic trend. This quantity represents the rate at which energy is dissipated per unit of granular temperature at a given driving frequency.
Its monotonic increase tells us that maintaining a stationary state with a granular temperature $K$ at higher driving frequencies implies a higher dissipation rate than at lower frequencies. In other words, our analysis provides a direct confirmation of our main argument for explaining the non-monotonic trend of the granular temperature as a function of $f$ discussed in the main text. Increasing the driving frequency increases the amount of energy available for absorption by the granular medium (i.e. the walls have a higher kinetic energy). However, it also increases the dissipation rate, which makes the energy absorption process less efficient. In the inset of Fig.~\ref{fig:Ediss} we also plot $\bar{\omega}_d$ as a function of $f$ showing that it is decreasing at large $f$. This tells us that the monotonic increase in ${\bar\omega}_d/K$ is due to the fact that the decay of dissipative power at large $f$ is slower than that of the granular temperature.

\section{A minimal two-block model for the $\bar{A}^2/\bar{p}$ scaling}\label{app:model}

\begin{figure}[h!]
\centering
\includegraphics[width=0.15\textwidth]{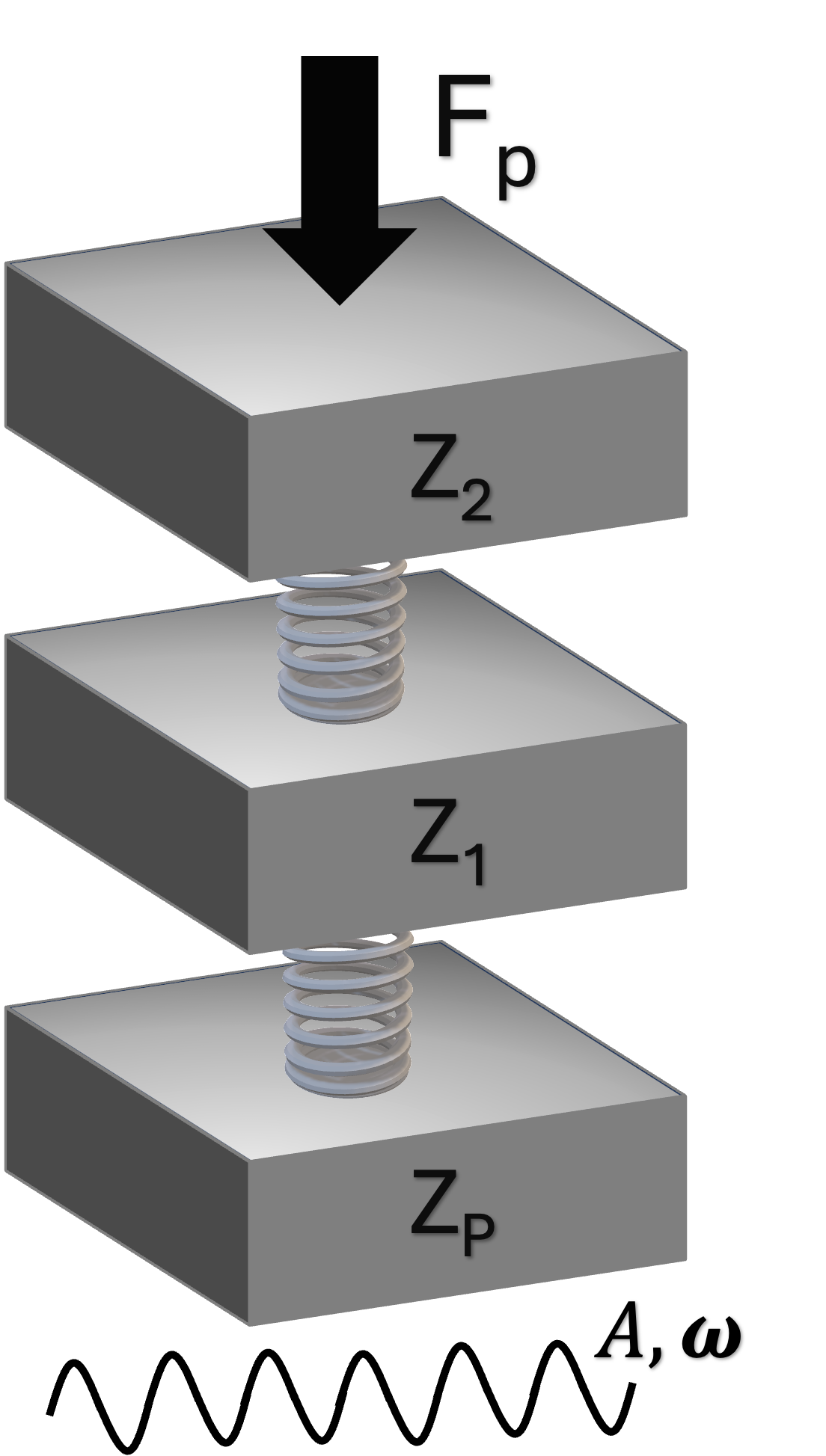}
\caption{Schematic representation of the two-block model.}
\label{fig:schematic}
\end{figure}

Here we present a generalized model of a driven-damped oscillator that qualitatively reproduces the main features of the frictional behavior observed in this study. The model is sketched in Fig.~\ref{fig:schematic}.

It extends the model introduced in~\cite{Plati2021Getting} to account for the kinetic energy measured in DEM simulations, which shows a non-monotonic dependence on the driving frequency with a maximum at $f^*$, an overall increase with driving amplitude, and a shift of the peak frequency toward lower values as dissipation grows.  

The model is derived from the equation of motion of a single particle, {by} assuming that the collective dynamics of the granular system can be approximated by the behavior of an individual grain. Since the medium remains dense across all shaking regimes, particle motion at short timescales is confined within cages formed by surrounding grains. The key parameter is the mean collision time $\tau_c$, which decreases as the driving frequency increases, leading to more frequent dissipative events.  

To keep the model simple, we restrict the analysis to vertical motion and describe a particle confined in a one-dimensional cage, represented by a spring attached to the base of a vibrating box oscillating with amplitude $A$ and frequency $f$. The cage, composed of neighboring grains, also induces an effective viscosity. Numerical simulations show that $\tau_c$ becomes smaller for increasing the driving frequency $f$. This effect has been modeled by assuming a viscosity $\gamma$ which is an increasing function of $f$: $\gamma=\gamma_0 f^{\alpha}$ with $\alpha\simeq 0.67$ estimated from numerical simulations.

To study the effect of friction, we consider a similar model in which a second particle is placed on top of the system to mimic the role of the upper plate over the granular material. A repulsive spring of stiffness $k_2$ is then included in the model to describe the interaction between the grains and the top plate.

Accordingly, because of the interaction with the top plate, the particle  is subject to a spring-like force of the form  
\begin{equation}
F_{\text{spring}} = 
%-k_p \, \left| z - z_p \right|^{\beta} %\, \text{sign}(z - z_p)
- k_2 \, \left| z - z_t \right|^{\beta} \, \text{sign}(z - z_t),
\end{equation}
where $z$ is the position of the particle and $z_t$ the position of the top plate. 
Here $\beta = 1$ or $\beta = 3/2$, depending on whether a Hookean or a Hertzian spring model is considered, respectively.  A similar force acts on the particle because of the interaction via a spring of stiffness $k_1$ with the bottom plate, whose position $z_p$ presents regular oscillations $z_p(t) = A \cos(2\pi f t)$.

In addition, the particle experiences a viscous force 
\begin{equation}
F_{\text{viscous}} = -\gamma \, (\dot{z} - \dot{z}_p),
\end{equation}
where $\gamma$ is the viscous damping constant.

The top plate is subject to two forces: an upward force from the interaction with the particle,
\begin{equation}
f_b = k_2 \, \left| z - z_t \right|^{\beta} \, \text{sign}(z - z_t),
\end{equation}
and a downward force $F_P$ proportional to the external pressure.  For simplicity, we consider a unitary mass $m=1$ for both the particle and the top plate.

The key quantity is the difference between the top plate position $z_t$ and the particle position $z$. When $z_t - z > l_0$, with $l_0$ the rest length of the spring, there is no interaction between the top plate and the grains. This corresponds to a zero-friction scenario. Conversely, friction arises when the top plate is ``attached'' to the particle.  

The mean value of friction coefficient can thus be estimated as the percentage of contact time:
\begin{equation}
\langle \mu \rangle = \frac{\tau_{\text{contact}}}{\tau_{\text{tot}}},
\end{equation}
where $\tau_{\text{tot}}$ is the total observation time and $\tau_{\text{contact}} \le \tau_{\text{tot}}$ is the fraction of time with $z_t - z \le l_0$.  
When the two particles are always in contact, $\langle \mu \rangle$ reaches its maximum value, while $\langle \mu \rangle = 0$ corresponds to the absence of contact.

To evaluate $\langle \mu \rangle$, it is necessary to solve the coupled equations for the dynamics of $z(t)$ and $z_t(t)$. This is a non-trivial task even in the Hookean case, and therefore we present below results of the numerical solution of the equations.

\subsection{Results}

\begin{figure}[h!]
  \centering
  \includegraphics[width=0.99\columnwidth]{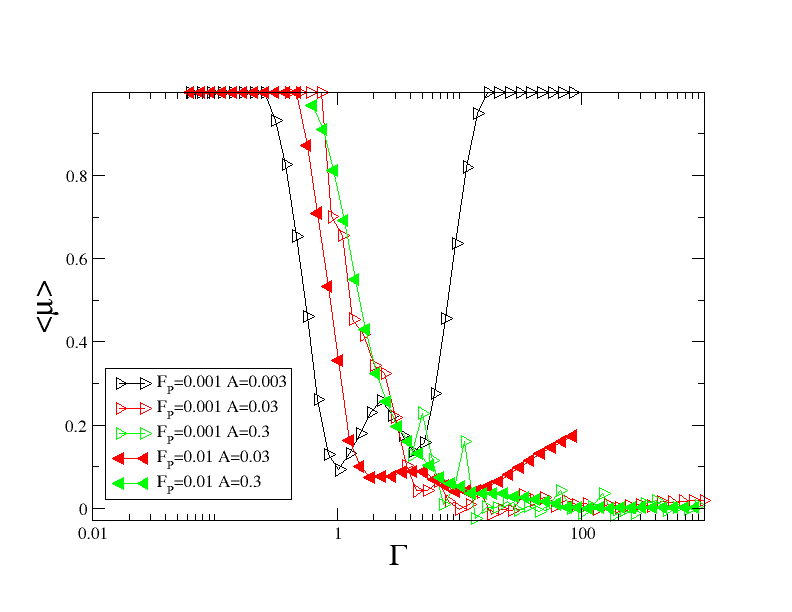}
  \caption{Average friction $\langle \mu \rangle$ as a function of 
  $\Gamma = A f^2/F_P$ in the Hertzian case. Different curves correspond to different values of $A$ and $F_P$. 
  The equations of motion are solved with $k_1=k_2=1$, and $\gamma=5$.}
  \label{frict_vs_gamma}
\end{figure}

In Fig.~\ref{frict_vs_gamma} we show the average friction $\langle \mu \rangle$ as a function of the dimensionless parameter $\Gamma = \frac{A f^2}{F_P}$ varies through $f$. 
Each curve corresponds to fixed values of $A$ and $F_P$, while the frequency $f$ is varied.  
Results are reported for the Hertzian case, but qualitatively similar behavior is found for the Hookean case.  

We observe a clear non-monotonic trend: friction decreases with increasing $\Gamma$, reaches a minimum $\mu_{\min}$ at $\Gamma > 1$, and then increases again. Both the position of the minimum and the value of $\mu_{\min}$ depend explicitly on $A$ and $F_P$.

\begin{figure}[h!]
  \centering
  \includegraphics[width=0.99\columnwidth]{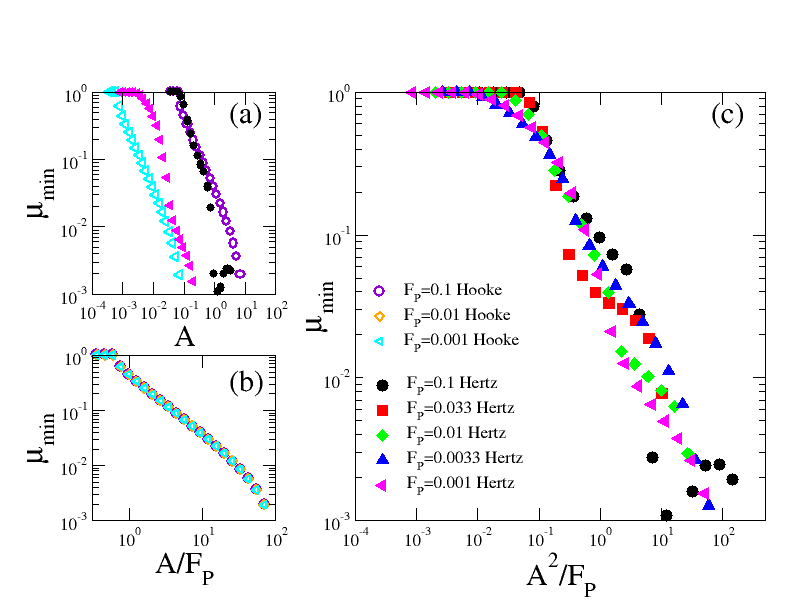}
  \caption{Analysis of $\mu_{\min}$ as a function of deformation amplitude and pressure. (a) Trend of $\mu_{\min}$ as a function of deformation amplitude ($A$) for two extreme pressure values: $F_P = 0.1$ (circles) and $F_P = 0.001$ (left triangles). Filled symbols represent the Hertzian contact model, while empty symbols represent the Hookean contact model. (b) Normalization of the Hookean data, showing the collapse of curves for various pressures as a function $A/F_P$. (c) Normalization of the Hertzian data, showing the collapse of curves as a function of $A^2/F_P$.}
  \label{mu_min}
\end{figure}

In Fig.~\ref{mu_min} we analyze the dependence of the minimum friction $\mu_{\min}$ on the driving amplitude $A$, for different confining pressures $F_P$, {by} comparing Hertzian and Hookean models.  
As shown in Fig.~\ref{mu_min}a, at small amplitudes $\mu_{\min} \simeq 1$, while for $A$ larger than a characteristic value (which increases with $F_P$), $\mu_{\min}$ decreases monotonically.  

A striking difference emerges between the two contact models.  
For the Hookean case (Fig.~\ref{mu_min}b), rescaling $A$ by $F_P$ leads to a perfect collapse of the data for different $F_P$.  
Conversely, in the Hertzian case (Fig.~\ref{mu_min}c), the collapse is recovered only when $A$ is rescaled by $F_P^{1/2}$.  

\subsection{Analytical interpretation of the different collapse of $\mu_{min}$ between Hookean and Hertzian cases}

Figure~\ref{mu_min} highlights the different scaling behaviors of the minimum friction $\mu_{\min}$ as a function of $A$ and $F_P$ for Hookean and Hertzian interactions.  
In the Hookean case, the scaling is exact: {by} rescaling $A$ and $F_P$ by the same factor yields identical results, i.e., the data are strictly invariant under proportional changes of $A$ and $F_P$.  
In contrast, for the Hertzian case the collapse obtained with $A/F_P^2$ is only approximate and limited to a certain range of parameters, with deviations emerging outside this regime.  
The origin of this difference can be clarified by analyzing a simplified version of the model, where the stiffness of the bottom spring is taken in the limit \( k_1 \to \infty \).  In this regime, the vertical position \( z_1 \) of the lower block is constrained to follow the bottom plate, i.e. \( z_1 = z_p + \ell_0 \), where \( z_p \) is the vertical position of the bottom plate and \( \ell_0 \) is the rest length of the lower spring.  
Thus, the dynamics reduce to those of the top block, whose vertical position \( z_2 \) satisfies the equation  

\begin{equation}
m \ddot{z}_2 + \gamma \dot{z}_2 + k_2 \left( \vert z_2 - z_1 \vert \right)^{\beta} \, \mathrm{sign}(z_2 - z_1) = N - F_P,
\label{x1}
\end{equation}
where \( N = k_2 \ell_0^{\beta} \) represents the reaction force ensuring that, in the absence of confining pressure (\( F_P = 0 \)), the equilibrium position of the top plate is \( z_2 = z_1 + \ell_0 \).

{By} assuming that the bottom plate oscillates as \( z_p = A \cos(\omega t) \), and {by} introducing the relative distance between the top block and the bottom plate as \( z = z_2 - z_p \), Eq.~(\ref{x1}) can be rewritten as  
\begin{equation}
\begin{split}
m \ddot{z} + \gamma \dot{z} + k_2 \left( \vert z - \ell_0 \vert \right)^{\beta} 
\, \mathrm{sign}(z - \ell_0)
= N - F_P \\
+\, m A \omega^2 \cos(\omega t) + \gamma A \omega \sin(\omega t).
\end{split}
\label{x2}
\end{equation}

{By} introducing dimensionless variables  
\[
y = \frac{z}{\ell_0}, \quad
A' = \frac{A}{\ell_0}, \quad
F'_P = \frac{F_P}{N}, \quad
\omega' = \omega t_0, \quad
\gamma' = \frac{\gamma \ell_0}{N t_0},
\]
with the characteristic time scale \( t_0 = \sqrt{\frac{m \ell_0}{N}} \),  
Eq.~(\ref{x2}) can be expressed in non-dimensional form as  

\begin{equation}
\ddot{y} + \gamma' \dot{y} + (1 + y)^{\beta} \, \mathrm{sign}(1 + y)
= 1 - F'_P + A' f(\omega'),
\label{y2}
\end{equation}
where derivatives are taken with respect to the scaled time \( t' = t / t_0 \), and  

\[
f(\omega') = \omega'^2 \cos(\omega' t') + \gamma' \omega' \sin(\omega' t').
\]

\subsubsection{Hooke case $\beta=1$}

{By} setting, $\beta=1$ in Eq.(\ref{y2}), after rescaling the variable as \( \upsilon = y / F'_P \), we obtain

\begin{equation}
\ddot{\upsilon} + \gamma' \dot{\upsilon} + \upsilon = 1 + \frac{A}{F'_P} f(\omega'),
\label{y3}
\end{equation}
which clearly shows that the evolution of \( \upsilon \) depends only on the ratio \( A / F'_P \).  
This implies that, upon rescaling the driving amplitude \( A \) by the external pressure \( F_P \), the explicit dependence on \( F_P \) vanishes.  
Such a scaling argument explains the exact data collapse observed in Fig.~\ref{mu_min}(b).

\subsubsection{Hertz case $\beta=3/2$}
In this case, we focus on small oscillations around the steady-state solution, i.e. \( y \ll 1 \).  
{By} expanding Eq.~(\ref{y2}) up to second order in \( y \), we obtain
\begin{equation}
\ddot{y} + \gamma' \dot{y} + \beta y + \lambda y^2 = -F'_p + A' f(\omega'),
\label{y3}
\end{equation}
with \( \lambda = 3/8 \).

We next decompose \( y(t) = q(t) + \psi(t) \), where \( q(t) = \langle y(t) \rangle \) denotes the average of \( y(t) \) over one period \( T = 2\pi / \omega \), and \( \psi(t) \) represents the fast fluctuating component.  
{By} substituting $y(t)$ into Eq.~(\ref{y3}) and averaging over one period yields
\begin{equation}
\ddot{q} + \gamma' \dot{q} + \beta q (1 + q/4) + \lambda \langle \psi^2 \rangle = -F'_p.
\label{q}
\end{equation}

The evolution equation for \( \psi(t) \) can be obtained by subtracting Eq.~(\ref{q}) from Eq.~(\ref{y3}).  
{By} following the so-called inertial approximation~\cite{Ble00}, we approximate it as
\begin{equation}
\ddot{\psi} = A' f(\omega'),
\label{psi}
\end{equation}
which gives \( \langle \psi^2 \rangle = \frac{\omega'^2}{2} \left( A'^2 \omega'^2 + \gamma'^2 \right) \).

{By} substituting this result into Eq.~(\ref{q}), we obtain
\begin{equation}
\ddot{q} + \gamma' \dot{q} + \beta q (1 + q/4)
= F'_P - \lambda \frac{\omega'^2}{2} \left( A'^2 \omega'^2 + \gamma'^2 \right).
\label{q2}
\end{equation}

Since this equation holds for \( y \ll 1 \), we introduce the rescaled variable \( \upsilon = q / F'_P \).  
At leading order, we obtain the oscillatory equation
\begin{equation}
\ddot{\upsilon} + \gamma' \dot{\upsilon} + \beta \upsilon
= 1 - \lambda \frac{ A'^2 \omega'^4}{2 F'_P}
  - \lambda \frac{\gamma'^2 \omega'^2}{F'_P}.
\label{q3}
\end{equation}

When \( A \omega^2 \) (the acceleration associated with the oscillating bottom plate) is much larger than \( \gamma \omega \) (the dissipative contribution of the medium), the oscillations of the upper block are primarily governed by the ratio \( A^2 / F_P \).  
This condition is expected to hold near the frequency corresponding to the minimum in the friction coefficient.  
Therefore, Eq.~(\ref{q3}) provides a theoretical justification for the scaling observed in Fig.~\ref{mu_min}c.  
In summary, our analysis demonstrates that the \( A^2 / F_P \) scaling naturally arises from non-linear effects in the effective interaction between the granular layer and the upper plate.

\end{document}